\documentclass[prl,twocolumn]{revtex4-1} 
\usepackage{natbib}

\usepackage{verbatim}
\usepackage{graphicx}
\usepackage[space]{grffile}
\usepackage{latexsym}
\usepackage{amsfonts,amsmath,amssymb}
\usepackage{url}
\usepackage[utf8]{inputenc}
\usepackage{fancyref}
\usepackage {cases}
\usepackage{float} 

\usepackage{ulem} 
\usepackage{color} 

\begin{document}

\title{Universality and Non-Universality in Distributed Nuclear Burning in Homogeneous Isotropic Turbulence}


\author{Yossef Zenati}%
 \email{ziyossef@campus.technion.ac.il}
\affiliation{%
 Physics Department, Technion - Israel Institute of Technology\\
 Haifa 3200003, Israel
}%

\author{Robert T. Fisher}%
 \email{robert.fisher@umassd.edu}
\affiliation{%
 University of Massachusetts Dartmouth, Department of Physics\\
 285 Old Westport Road, North Dartmouth, Ma. 02747 
}%

\date{\today}

\begin{abstract}
Nuclear burning plays a key role in a wide range of astrophysical stellar transients, including thermonuclear, pair instability, and core collapse supernovae, as well as kilonovae and collapsars. Turbulence is now  understood to also play a key role in these astrophysical transients. Here we demonstrate that turbulent nuclear burning may lead to large enhancements above the uniform background  burning rate, since turbulent dissipation gives rise to temperature fluctuations, and in general the nuclear burning rates are highly sensitive to temperature. We derive results for the turbulent enhancement of the nuclear burning rate under the influence of strong turbulence in the distributed burning regime in homogeneous isotropic turbulence, using probability distribution function (PDF) methods. We  demonstrate that the turbulent enhancement obeys a universal scaling law in the limit of weak turbulence. We further demonstrate that, for a wide range of key nuclear reactions, such as C$^{12}$(O$^{16}$, $\alpha$)Mg$^{24}$ and triple-$\alpha$, even relatively modest temperature fluctuations, of the order ten percent, can lead to enhancements of 1 - 3 orders of magnitude in the turbulent nuclear burning rate. We verify the predicted turbulent enhancement directly against numerical simulations, and find very good agreement. We also present an estimation for the onset of turbulent detonation initiation, and discuss implications of our results for the modeling of stellar transients.  
\end{abstract}


\maketitle

\textit {Introduction.} Nuclear energy plays a fundamental role in stellar astrophysics, providing the ultimate power source of all stars \citep {eddington26, b2fh}. In particular, nuclear reactions play an important role in a wide range of stellar astrophysical transients, powering thermonuclear and pair instability supernovae, liberating the neutrinos which give rise to core collapse supernovae, and synthesizing r-process heavy nuclei in kilonovae and collapsars \cite {aspdenetal08, couchott15, radice17}.

The inferred Reynolds numbers of stellar transients are typically extremely large, implying that their stellar plasmas are highly turbulent \citep{brandenburgnordlund11, fisheretal19}. Turbulence naturally gives rise to stochastic fluctuations in velocity as well as in temperature and density, with important consequences for nuclear burning. In particular, because of the strong energy dependence of the quantum nuclear penetration factor, thermonuclear reaction rates are extremely sensitive to temperature, and modest thermodynamic fluctuations enhance their rates dramatically. Previous authors have used theory to examine how small-scale turbulent intermittent  velocity fluctuations may influence the burning rate  \citep {lisewskietal00a, lisewskietal00b, panetal08}. In this Letter, we focus specifically upon the role which the intrinsically stochastic turbulent thermodynamic fluctuations have upon key thermonuclear rates, using a combination of both theory and three-dimensional hydrodynamical simulations.

\textit {Analytic Derivation of Turbulent Nuclear Burning Enhancement for Power-Law Rates}. When the turbulent background is weak, and the fuel and ash are spatially separated, the burning proceeds through a thin laminar flame surface whose thickness is established by the balance of nuclear energy release and thermal conduction. In contrast, when turbulence is sufficiently strong, the flame surface is completely disrupted, causing burning to develop throughout the volume in the distributed nuclear burning regime \citep {poinsotveynante}. The dimensionless Karlovitz number, defined as ${\rm Ka} = \sqrt {u^3 l / (s_l^3 L)}$, determines the relative importance of turbulence in a reactive medium. Here $u$ is the root-mean-squared (RMS) velocity on the integral scale $L$, and $s_l$ and $l$ are the laminar flame speed and thickness, respectively \citep {aspdenetal08}. When Ka $< 1$, turbulence plays a minor role on the scale of the flame, and the flame remains laminar. For large Ka $\gg 1$, the flame is disrupted by the turbulence, and exists in the distributed burning regime \citep {aspdenetal08}. This distributed burning regime is most relevant to strongly dynamical stellar transients, arising for instance in accretion flows in white dwarf mergers \citep {fisheretal19} and in X-ray bursts on neutron stars \citep {timmesniemeyer00}. In this Letter, we focus on distributed nuclear burning, and assume that the burning is sufficiently weak that its backreaction upon the turbulent velocity and temperature fields can be neglected. These assumptions apply up to the onset of detonation initiation \citep {fisheretal19}. Under our presumed conditions, temperature, density, and abundances all behave as passive scalars. 

 
The power spectrum of a scalar field is defined to be the angle-averaged Fourier transform of its spatially-averaged autocorrelation function.  However, unlike the statistics of the velocity field, passive scalars in general depend upon initial and boundary conditions \citep [see for example,][] {warhaft00}. The inertial range scaling of temperature fluctuations in homogeneous and isotropic turbulence was first argued by Obukhov and Corrsin \citep {obukhov49, corrsin51}, and later verified experimentally  \citep {warhaft94, niemelaetal00}, to have Gaussian statistics with the same power law index as the Kolmogorov velocity spectrum  :


\begin {equation}
E_T (k) \propto k^{-5/3}
\end {equation}
%

We next consider the question of the scale dependence of the burning rate. The integral of the power spectrum is, by Parseval's theorem, the square of the RMS temperature fluctuation over the spatial domain, ${\delta T}$ :
\begin {equation}
{\delta T}^2 = \int E_T (k) dk
\end {equation}
Consequently, Obhukov-Corrsin scaling implies the RMS temperature fluctuation on a length scale $r$, $\delta T (r)$, similarly follows a Kolmogorov-like distribution,
\begin {equation}
\label {eq:trms}
\delta T (r) = \delta T \left( {r \over L} \right)^{1/3}
\end {equation}
The Obukhov-Corrsin scaling of turbulent temperature fluctuations neglects the influence of turbulent intermittency, which can play an important role on small scales much less than the integral scale $r \ll L$ \citep {benzietal92}.  However, just as the turbulent specific kinetic energy ${\delta v (r)}^2$ is greater on larger length scales, so too is the temperature fluctuation ${\delta T (r)}^2$ greater on large length scales. Consequently, the largest length scales, comparable to the integral scale, must dominate the turbulent nuclear burning in the distributed burning regime. Here we focus upon the influence of turbulence on these largest scales in homogeneous and isotropic turbulence, whose passive scalars are well-described by Obukhov-Corrsin scaling. 


We express the local specific energy generation rate at a given point as $\epsilon (X, \rho, T)$. We model the specific energy generation rate as the power-law expression $\epsilon (X, \rho, T) = \epsilon_0 X^{m+1} \rho^m T^n$, where $X$ is the mass fraction, $\rho$ is the density, for a $(m + 1)$-body single species reaction. The results obtained can be easily extended to multiple species reactions. We express the mean temperature as $T_0$, the mean density as $\rho_0$, and the mean abundance as $X_0$, averaged over a spherical volume of radius $r$. 

Integrating the energy generation weighted by the joint probability distribution function (PDF) of composition, density, and temperature, $P_r (X, \rho, T)$, we obtain the volume-averaged energy generation rate $\epsilon_r$ on the length scale $r$:

%

\begin {multline}
\epsilon_r  ({\delta X / X_0}, {\delta \rho / \rho_0} , {\delta T / T_0}) =\\  \int dX \int d\rho \int dT\ \epsilon (X, \rho, T) P_r (X, \rho, T)
\end {multline}
This volume-averaged energy generation rate plays a key role in computational hydrodynamical modeling of the nuclear burning rate, serving as a source term to the energy equation.

Let us assume the PDFs of temperature, density, and composition individually follow Gaussian distributions,

\begin {equation}
  P_r (T) = { 1 \over \sqrt {2 \pi \delta T(r)^2} } \exp {\left(-\frac {(T - T_0)^2} {2 \delta T(r)^2 }\right)},
\end {equation}
\begin {equation}
  P_r (\rho) = { 1 \over \sqrt {2 \pi \delta \rho(r)^2} } \exp {\left(-\frac {(\rho - \rho_0)^2} {2 \delta \rho(r)^2 }\right)},
\end {equation}
\begin {equation}
  P_r (X) = { 1 \over \sqrt {2 \pi \delta X(r)^2} } \exp {\left(-\frac {(X - X_0)^2} {2 \delta X(r)^2 }\right)}.
\end {equation}
The joint PDF $P_r (X, \rho, T)$ is then a multivariate Gaussian distribution, including possible correlations between density, temperature, and composition. We illustrate the calculation of this  integral by focusing on the density-temperature correlation only, before extending the result to the full multivariate distribution.

The joint bivariate distribution for density and temperature $\rho$ and $T$ with correlation $r_{\rm corr} (\rho, T)$ can be expressed in terms of uncorrelated normal variates $x$ and $y$:

\begin {equation}
T = \delta T (r) x + T_0
\label {temperature_def}
\end {equation}

\begin {equation}
\rho = \delta \rho (r) \left (r_{\rm corr} (\rho, T) x + \sqrt {1 - r_{\rm corr}^2 (\rho, T)} y  \right) + \rho_0
\label {density_def}
\end {equation}
Here the density-temperature correlation $r_{\rm corr} (\rho, T) $ is defined as usual as:

\begin {equation}
r_{\rm corr} (\rho, T) = {\iint d\rho dT P_r (\rho, T) \rho T \over {\delta \rho\ \delta T} }
\end {equation}
As can be expected, simulations of driven homogeneous isotropic turbulence exhibit a strong positive correlation of density and temperature. 
From the 3D simulations presented later in this Letter, we have computed  $r_{\rm corr} (\rho, T) \simeq 0.5$. 

The dimensionless normal variate $x$ has a simple interpretation. $x$ is the ratio of the differential temperature $T - T_0$ to the RMS temperature fluctuation on the scale $r$, $\delta T (r)$,
\begin {equation}
x = { {T  - T_0} \over \delta T (r) }  = {T - T_0 \over \delta T} \left (L \over r \right)^{1/3}
\end {equation}
Similarly, in the absence of correlations, when $r_{\rm corr} = 0$,  $y$ can be interpreted as the ratio of the differential density fluctuation $\rho - \rho_0$ to the RMS density fluctuation on the scale $r$, $\delta \rho (r)$.


The turbulent enhancement $\epsilon_r (\delta \rho / \rho_0, \delta T / T_0)$ is then expressed as an integral over the uncorrelated $x$ and $y$ variates,

\begin {multline}
\epsilon_r  ({\delta \rho / \rho_0} , {\delta T / T_0}) = \\
{1 \over 2 \pi} \int dx \int dy\ \epsilon (\rho, T) \exp \left [- {1 \over 2} \left (x^2 + y^2 \right)     \right],
\label {jointpdf_turb_integral}
\end {multline}
where we have used the fact that the joint PDF of the normal variates $x$ and $y$ are uncorrelated. Substituting equations  \ref {temperature_def} and \ref {density_def} into equation  \ref {jointpdf_turb_integral}, one can evaluate the resulting Gaussian integrals by writing $T^n = T_0^n \left [1 + \delta T / T_0 (r/L)^{1/3} x\right]^n$, and similarly for $\rho$, and then expanding these expressions using the binomial theorem. One finds the general turbulent enhancement for two-body power law rates ($m = 1$) including density-temperature correlations. A complete calculation extending this derivation for the full multivariate distribution, including all cross-correlations, yields

\begin {widetext}

\begin {equation}
\begin {split}
{\epsilon_r  ({\delta X / X_0}, {\delta \rho / \rho_0} , {\delta T / T_0})  \over \epsilon  (X_0, \rho_0, T_0)} = \sum_{\substack{k = 0 \\ k\ {\rm even} }}^n {n!  (k - 1)!!  \over (n - k)! \enspace k!} \left[1 + \left\{ 1 + k r_{\rm corr}^2 (X, T) \right\} \left(\delta X \over X_0 \right)^2 \left (r \over L   \right)^{2/3}    \right]  \left ({\delta T \over T_0} \right)^{k} \left( {r \over L} \right)^{k/3} \\ 
+  r_{\rm corr}  (\rho, T) \left ({\delta \rho \over \rho_0} \right) \sum_{\substack{k = 1 \\ k\ {\rm odd} }}^n {n!  k!!  \over (n - k)! \enspace k!} \left ({\delta T \over T_0} \right)^k \left( {r \over L} \right)^{(k + 1)/3} 
+ 2 r_{\rm corr}  (X, T) \left ({\delta X \over X_0} \right) \sum_{\substack{k = 1 \\ k\ {\rm odd} }}^n {n!  k!!  \over (n - k)! \enspace k!} \left ({\delta T \over T_0} \right)^k \left( {r \over L} \right)^{(k + 1)/3} \\
+ 2 r_{\rm corr} (X, \rho)  \left ({\delta X \over X_0} \right)  \left ({\delta \rho \over \rho_0} \right) \left( {r \over L} \right)^{2/3}
\end {split}
\label {eqn:fullenhance}
\end {equation}

\end {widetext}

The double factorial function  is defined as the factorial function including only those factors with the same parity (even or odd) as the argument.

In the weak enhancement regime (${\delta X / X_0} (r / L)^{1/3} \ll 1$, ${\delta T / T_0} (r / L)^{1/3} \ll 1$, ${\delta \rho / \rho_0} (r / L)^{1/3} \ll 1$), the turbulent enhancement of the averaged burning rate $\epsilon_r$ grows quadratically with temperature, density, and composition fluctuations. For two-body interactions ($m = 1$), this weak limit yields


\begin {equation}
\begin {split}
{\epsilon_r  ({\delta X / X_0}, {\delta \rho / \rho_0} , {\delta T / T_0})  \over \epsilon  (X_0, \rho_0, T_0)}  \simeq 1 +  
\Bigg[ {n (n - 1) \over 2} \left( {\delta T \over T_0}\right)^2  \\
+ n   \left( {\delta T \over T_0}\right) \left \{ r_{\rm corr}(\rho, T) \left( {\delta \rho \over \rho_0}\right)  + 2  r_{\rm corr}(X, T) \left( {\delta X \over X_0}\right) \right \}  \\
+ 2 r_{\rm corr} (X, \rho) \left( {\delta X \over X_0}\right)  \left( {\delta \rho \over \rho_0}\right)  
+  \left( {\delta X \over X_0} \right)^2 \Bigg]
\left ({r \over L }\right)^{2/3} 
\end {split}
\label {eqn:weakenhance}
\end {equation}


%
A similar calculation can be carried out for three-body interactions ($m = 2$), with the resulting enhancement also scaling as $r^{2/3}$, including all cross-correlations.

It can be seen that, regardless of the reaction, the amplitude of the turbulent fluctuations, and the correlation between composition, density, and temperature fluctuations, the enhancement of the averaged burning rate $\epsilon_r$ scales as $(r/L)^{2/3}$ in the weak enhancement regime. Crucially, {\it the dependence of the averaged burning rate upon length scale is universal for weak homogeneous, isotropic turbulence}. 
Physically, this universality of the turbulent enhancement of the averaged burning rate can be understood as a direct manifestation of the universality of the temperature field in Obukhov-Corrsin turbulence, eqn. \ref {eq:trms}.

\begin{figure}[h]
	\begin{center}
		\includegraphics [width=0.9\columnwidth]{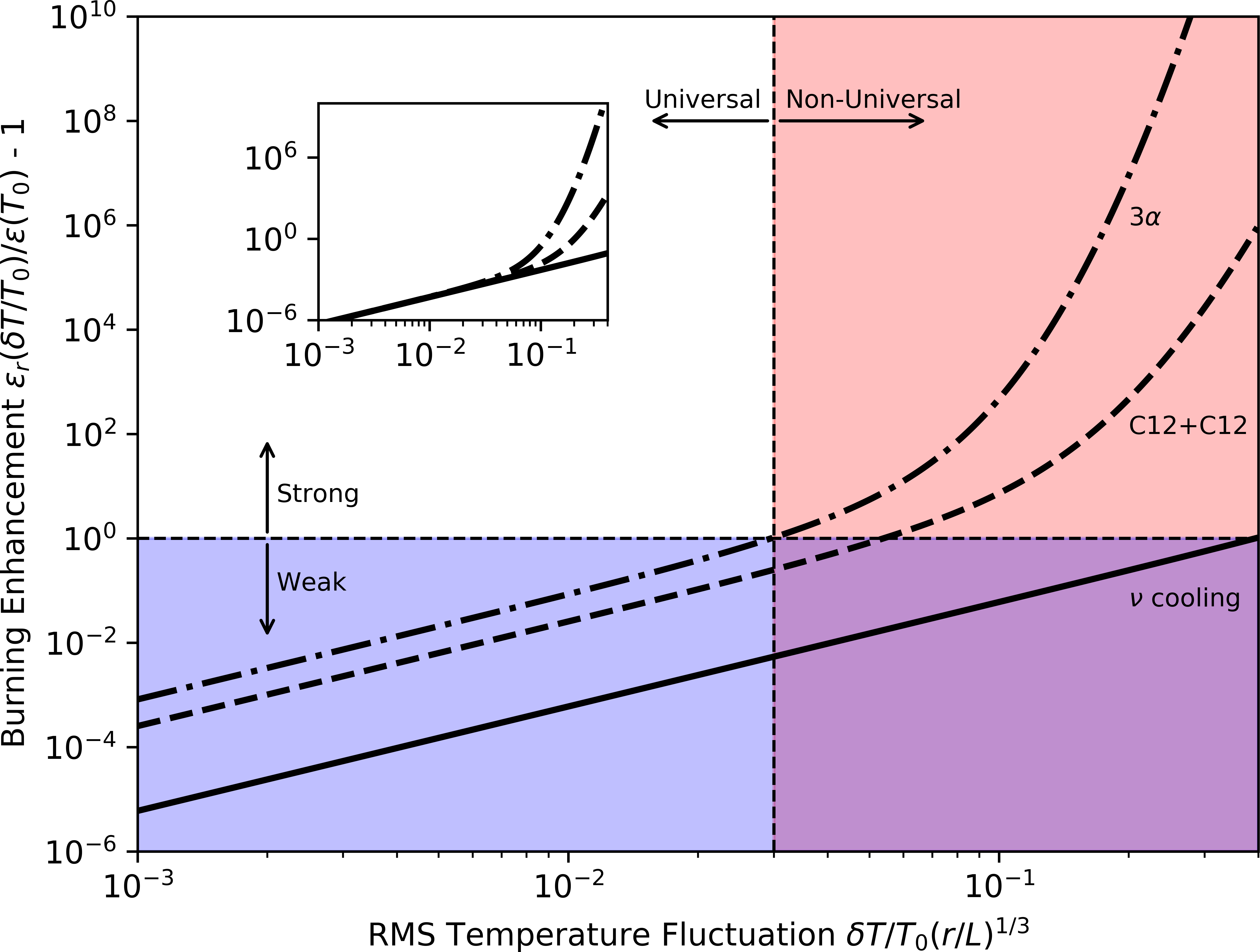}
		\caption{A log-log plot of the dimensionless fractional turbulent enhancement in the nuclear burning rate $\epsilon_r (\delta T / T_0) / \epsilon (T_0) - 1$, as a function of the RMS temperature fluctuation on length scale $r$, normalized to the mean temperature $T_0$, $ {\delta T / T_0} \left (r / L \right)^{1/3}$, in the distributed burning regime. The curves shown are for neutrino cooling via the URCA process (solid line, $n = 8$), $^{12}$C-$^{12}$C burning (dashed, $n = 23$), and triple-$\alpha$ reaction (dot dashed, $n = 41$). The inset figure shows the same three curves on the same set of axes, compensated by the factor $[n (n - 1) ]^{-1}$. For weak enhancement, the compensated enhancement collapses onto a single curve, demonstrating its universal nature.}
		\label {fig:turb_enhance}
	\end{center}
\end{figure}

For realistic nuclear burning rates which typically have $n \gg 1$, a useful simplified expression for the volume-averaged turbulent burning rate which includes only the temperature fluctuations is 

\begin {multline}
{\epsilon_r (\delta T / T_0) \over \epsilon (T_0)} = \int  {dx \over  \sqrt {2 \pi} } \left [1 + {\delta T \over T_0} \left (r \over L \right)^{1/3} x \right]^n \exp(-x^2 /2)\\
=  \sum_{\substack{k = 0 \\ k\ {\rm even} }}^n {n!  (k - 1)!!  \over (n - k)! \enspace k!} \left[ \left ({\delta T \over T_0} \right) \left( {r \over L} \right)^{1/3} \right]^k
\label {eqn:tempefluctuations}
\end {multline}
Because the fluctuations $(\delta T / T_0)$, $(\delta \rho / \rho_0)$, and $(\delta X / X_0)$ are of the same order, this expression is typically accurate to within a factor of $1/n$ or better ($2.5 - 5\%$ for a range of astrophysically-relevant reactions, with $n = 20 - 40$) of the complete result, eqn. \ref {eqn:fullenhance}, including all fluctuation terms and their correlations.  Figure \ref {fig:turb_enhance} plots the fractional enhancement $\epsilon_r (\delta T / T_0) / \epsilon (T_0) - 1$ as a function of the RMS temperature fluctuation on the the scale $r$, $(\delta T / T_0) (r / L)^{1/3}$ for several representative reactions, with the value of $n$ taken at relevant temperatures. This plot illustrates both the universal scaling of the enhancement for weak turbulence, as well as the non-universal enhancement for strong turbulence.

\textit {Verification of Turbulent Nuclear Burning Enhancement}. The analytic predictions of the preceding section can be verified by comparison against three-dimensional numerical simulations of turbulent distributed nuclear burning. For this purpose, we have used the FLASH4 code \citep {fryxell00}.  
We employ the Helmholtz equation of state, which incorporates both ions (treated as an ideal gas) as well as electrons with an arbitrary degree of degeneracy and special relativity \citep {Timmes_2000}. Nuclear burning is included using a 19- isotope network with 78 rates\citep {weaveretal78}, and optimized in a hardwired implementation\citep {timmes99}.

The simulations presented in this Letter are initially static and uniform, at a resolution of $256^3$, and are driven by a large-scale stochastic forcing routine \cite {fisheretal08, federrathetal10} with nuclear burning turned off. Once a steady-state is achieved in the RMS velocity and the enstrophy, nuclear burning is activated. The turbulence-driving methodology has been extensively verified and validated against both theory and experiment -- see \citet {fisheretal19} and references therein.

\begin{figure}[h]
	\begin{center}
		\includegraphics[width=0.9\columnwidth]{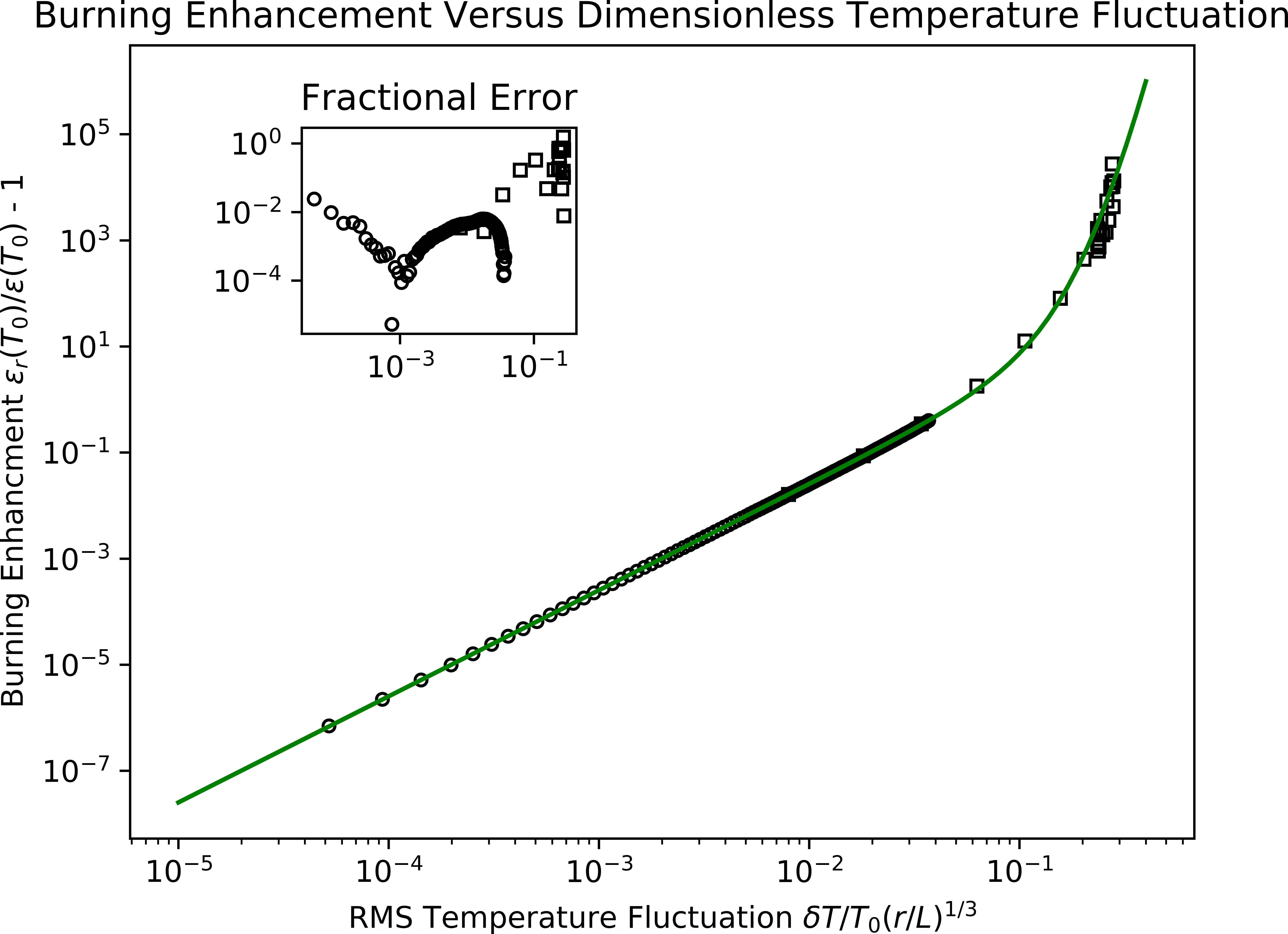}
		\caption{Figure showing the verification of the predicted turbulent burning rate enhancement. See text for description.}
		\label {fig:comparison_plot}
	\end{center}
\end{figure}

The results of the verification are shown in figure \ref {fig:comparison_plot}. A time series of the simulated power-law burning rate sampled from the turbulent simulations, calculated with $n = 23$, appropriate to C12 fusion in the regime of astrophysical interest, is plotted against the predicted rate. The  plot shows the predicted fractional turbulent enhancement compared with simulated turbulent data versus the dimensionless RMS temperature fluctuation $\delta T / T_0 (r / L)^{1/3}$, for two turbulent models, with $\delta T / T_0 = 0.03$ and $\delta T / T_0 = 0.3$, shown in the open circles and open squares, respectively. The inset shows the fractional log error versus the dimensionless temperature fluctuation.

The simulations are in very good agreement with the predicted enhancement rates throughout the weak regime ($\delta T / T_0 \ll 0.1$), with typical fractional errors much less than 1\%. Notably, this agreement applies even in the driving regime, when the turbulence is not yet in steady-state, because it is the large-scale temperature fluctuations that dominate the enhancement. Furthermore, the results also show good agreement, to within 10\%, into the moderately strong regime ($\delta T / T_0 \simeq 0.1$). It is only at $\delta T / T_0 > 0.1$ that the fractional errors become of order unity. At these higher levels of turbulent temperature fluctuations, higher-order moments of the temperature distribution become more important in the calculated enhancement rates, and possible departures from Gaussianity in the numerical simulations may become important. 

\textit {Turbulent Detonation Initiation}. With a description of turbulent enhancement of nuclear burning in place, we next address the conditions under which burning may transition to a detonation. Detonation arises during supersonic burning. Consequently, we construct a simple estimate for the the conditions for detonation to arise within a distributed burning region on scale $r$ by comparing the sound-crossing time to the nuclear burning timescale. The sound-crossing time on the scale $r$ is simply $\tau_{\rm sound} (r) = r / c_{\rm s}$. The nuclear burning timescale $\tau_{\rm nuc} (r)$ on the scale $r$, including the turbulent nuclear enhancement, is given by:

\begin {equation}
\tau_{\rm nuc} (r) =  {c_p T_0 \over n \epsilon_r (\delta T / T_0) \epsilon (T_0)}
\label {eqn:taunuc}
\end {equation}
Here $c_p$ is the ratio of specific heats at the background density and temperature $T_0$.  Our equation \ref {eqn:taunuc} is closely related to equation 18 from \citep {woosley07}. In his analysis, Woosley evaluates the nuclear burning time at some temperature in the isobarically-mixed ash, and reduces to our expression when $\epsilon_r (\delta T / T_0) = 1$. By including the turbulent enhancement due to temperature fluctuations consistently, we in general obtain a much shorter nuclear burning timescale, and consequently a wider range of conditions susceptible to detonation. 

\begin{figure}[h]
	\begin{center}
        \includegraphics[width=0.9\columnwidth]{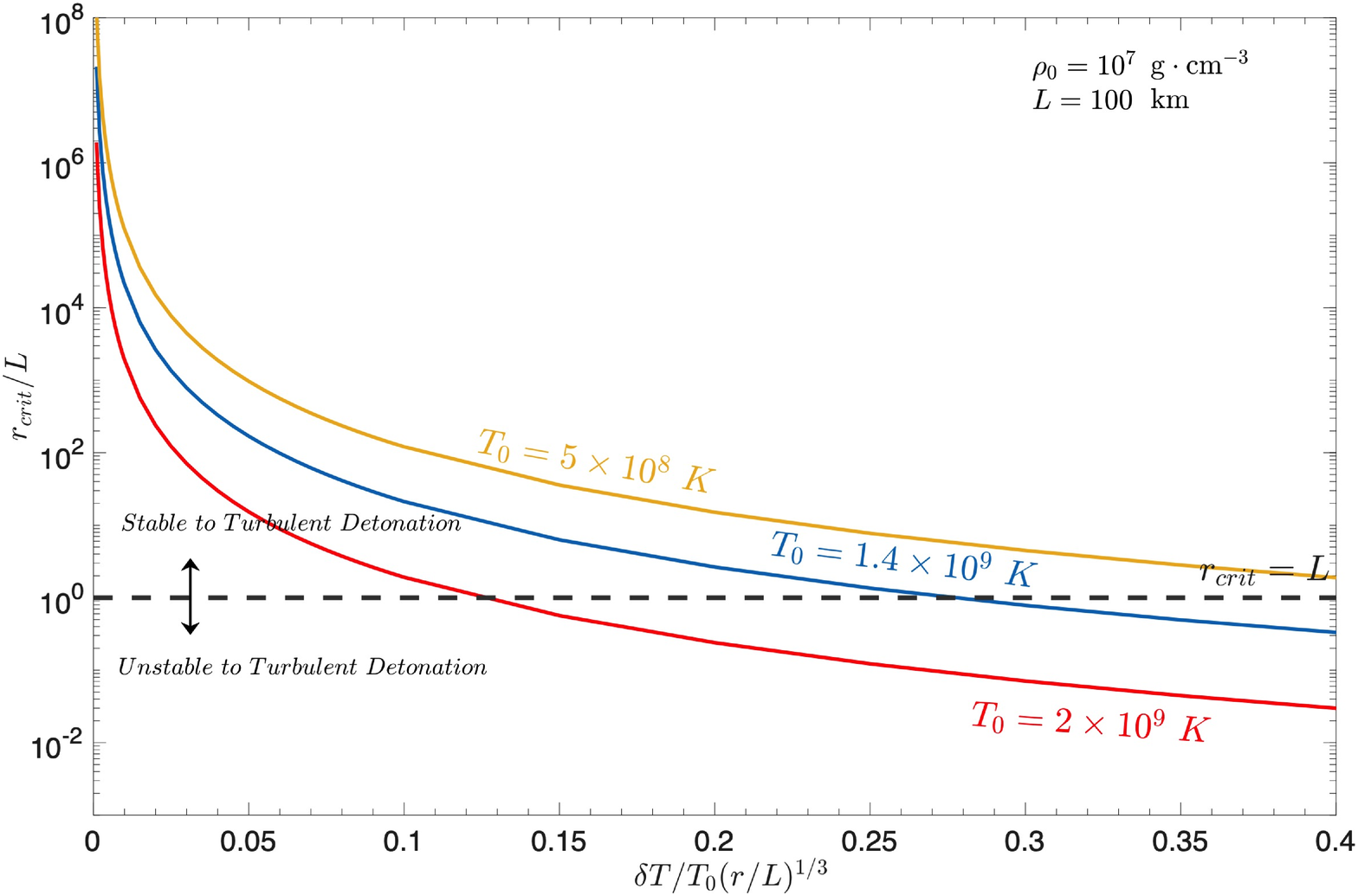}
        \caption{A log-linear plot of the ratio of the critical length to the integral scale,  $r / L$ versus the  temperature fluctuation $\delta T / T_0 (r / L)^{1/3}$ on the scale $r$. Above this curve, the conditions are stable to detonation, below the curve they are unstable to detonation.}
        \label {fig:strong_regime}
    \end{center}
\end{figure}

The condition that $\tau_{\rm nuc} (r) < \tau_{\rm sound} (r)$ is satisfied for length scales $r$ above a critical length scale $r_{\rm crit}$.  In figure \ref {fig:strong_regime}, we plot the ratio of the critical length scale $r_{\rm crit}$ to the integral scale $L$, as a function of the temperature fluctuation $\delta T / T_0 (r / L)^{1/3}$ on the scale $r$, for  C$^{12}$(O$^{16}$, $\alpha$)Mg$^{24}$.  
The background density and integral scale are held fixed at $\rho_{0}=1\times 10^{7}$\ g cm$^{-3}$ and $L = 100$ km, while the background temperature for three representative cases: $T_0 = 5 \times 10^8$ K, $1.4 \times 10^9$ K, and $2 \times 10^9$ K. 

In figure \ref {fig:strong_regime}, the horizontal dashed line demarcates the critical threshold of $r_{\rm crit} / L = 1$. Above this line, the critical length scale is larger than the integral scale $L$, and the flow experiences stable distributed nuclear burning. Below this line, the critical length scale is smaller than the integral scale, and the distributed burning regime becomes unstable to detonation initiation. Our simple estimate predicts temperature fluctuations of order 10\% are sufficient to produce a detonation upon a background temperature of $T_0 = 2 \times 10^9$ K, with increasingly stronger temperature fluctuations required for colder temperature backgrounds. These findings, based upon the single  C$^{12}$(O$^{16}$, $\alpha$)Mg$^{24}$ reaction, are in rough agreement with a series of detailed numerical simulations \citep {fisheretal19} with a full reaction network, where it was demonstrated that 10\% temperature fluctuations on a $T_0 = 1.2 \times 10^9$ K background led to detonation initiation.

\textit {Discussion.} The $r^{2/3}$ scaling of the turbulent enhancement of the distributed burning regime found in eqn. \ref {eqn:weakenhance} has been discussed by other authors. For instance \citet {aspdenetal11}, derive it by considering the turbulent flame speed, assuming that the turbulent burning time scale is constant.  Crucially, our approach clarifies that the $r^{2/3}$ scaling is an approximation which applies in the limit of weak enhancement in Obhukov-Corrsin turbulence only. In fact, for even modestly strong turbulent enhancement, the scaling behavior can be greatly different, with increasingly stronger sensitivity to the scale length $r$ for increasing turbulent fluctuations. 

The results obtained for turbulent enhancement and detonation initiation may be incorporated into three-dimensional subgrid models models for nuclear burning within stellar transients. As we have demonstrated, an accurate determination of strong turbulent enhancement as calculated directly from numerical simulations requires high-ordered moments of the temperature distribution -- e.g. the 8th-13th moments for $n \simeq 20 - 40$. An accurate determination of such high-ordered moments in turn requires large statistical samples, which in turn necessitates large numbers of cells within an integral scale. The challenge of modeling stellar transients in full three dimensions generally means that the integral scale of turbulence is only very coarsely resolved, if at all. In contrast, the analytic calculation of the turbulent enhancement for Gaussian statistics requires only the RMS temperature fluctuation, which can be estimated accurately with far fewer cells. Consequently, the formalism developed here provides a promising basis for an approach for subgrid modeling of turbulent nuclear burning and detonation initiation within the distributed burning regime in three-dimensional simulations of stellar transients.




\begin{acknowledgements} 
The authors thank Prof. Hagai B. Perets and Evgeni Grishin for stimulating discussions. R.T.F. thanks the Institute for Theory and Computation at the Harvard-Smithsonian Center for Astrophysics, and the Kavli Institute for Theoretical Physics, supported in part by the national Science Foundation under grant NSF PHY11-25915, for visiting support during which a portion of this work was completed. R.T.F. acknowledges support from NASA 80NSSC18K1013. This work used the Extreme Science and Engineering Discovery Environment (XSEDE) Stampede 2 supercomputer at the University of Texas at Austin’s Texas Advanced Computing Center through allocation TG-AST100038, supported by National Science Foundation grant number ACI-1548562.

We use a modified version of the FLASH code 4.0, which was in part developed by the DOE NNSA-ASC OASCR Flash Center at the University of Chicago \citep {fryxell00}. Our analysis and plots strongly benefited from the use of the yt package \citep {turk_2011}.
\end {acknowledgements}


\bibliographystyle{apsrev}
\bibliography{distributed_burning}


\end{document}